\documentclass[aps,prb,showpacs,twocolumn,superscriptaddress]{revtex4-2}

\usepackage{microtype}	% Besserer Blocksatz

\usepackage{amsmath}
\usepackage{amsfonts}
\usepackage{amssymb}

\usepackage{braket}
\usepackage{tensor}

\usepackage{graphicx}
\usepackage{pgfplots}
\pgfplotsset{compat=1.17}

\usepackage{xcolor}

\definecolor{faublue}{rgb}{0,0.22,0.40}				%003865
\definecolor{fauyellow}{rgb}{0.79,0.58,0.07}		%c99313
\definecolor{faugrey}{rgb}{0.60,0.64,0.68}			%98a4ae
\definecolor{faugreen}{rgb}{0,0.61,0.47}			%009b77
\definecolor{faulightblue}{rgb}{0,0.69,0.92}		%00b0eb
\definecolor{faured}{rgb}{0.55,0.08,0.16}			%8D1429
\usepackage[colorlinks=true, linkcolor=faublue, citecolor=faublue, urlcolor=faublue]{hyperref}

\usepackage{array}
\newcolumntype{C}[1]{>{\centering\arraybackslash}m{#1}}

\newcommand{\imag}{\mathrm{i}} 
\newcommand{\sgn}[1]{\mathrm{sgn}(#1)}
\newcommand{\PT}{\mathcal{PT}}
\newcommand{\HInt}{\mathcal{H}_\textrm{I}}
\newcommand{\ic}{\mathcal{H}_\textrm{TFIM}}
\newcommand{\tc}{\mathcal{H}_\textrm{TCF}}
\newcommand{\tcg}{\mathcal{\widetilde{H}}_\textrm{TCF}}
\newcommand{\crit}{x_\text{c}}
\newcommand{\rc}{r_\text{c}}

\newcommand{\lf}{\textrm{lf}}
\newcommand{\hf}{\textrm{hf}}

\newcommand{\diff}{\mathop{}\!\mathrm{d}}

\usepackage[super]{nth}		%for ordinal numbers (1st, 2nd, 3rd, ...)

\begin{document}
		
	\title{High-order series expansion of non-Hermitian quantum spin models}	
	\author{Lea Lenke}
	\affiliation{Lehrstuhl f\"ur Theoretische Physik I, Staudtstra{\ss}e 7, Universit\"at Erlangen-N\"urnberg, D-91058 Erlangen, Germany}
	
	\author{Matthias M\"uhlhauser}
	\affiliation{Lehrstuhl f\"ur Theoretische Physik I, Staudtstra{\ss}e 7, Universit\"at Erlangen-N\"urnberg, D-91058 Erlangen, Germany}
	
	\author{Kai Phillip Schmidt}
	\affiliation{Lehrstuhl f\"ur Theoretische Physik I, Staudtstra{\ss}e 7, Universit\"at Erlangen-N\"urnberg, D-91058 Erlangen, Germany}
	
	\begin{abstract}
	 We investigate the low-energy physics of non-Hermitian quantum spin models with $PT$-symmetry. To this end we consider the one-dimensional Ising chain and the two-dimensional toric code in a non-Hermitian staggered field. For both systems dual descriptions in terms of non-Hermitian staggered Ising interactions in a conventional transverse field exist. We perform high-order series expansions about the high- and low-field limit for both systems to determine the ground-state energy per site and the one-particle gap. The one-dimensional non-Hermitian Ising chain is known to be exactly solvable. Its ground-state phase diagram consists of second-order quantum phase transitions, which can be characterized by logarithmic singularities of the second derivative of the ground-state energy and, in the symmetry-broken phase, the gap closing of the low-field gap. In contrast, the gap closing from the high-field phase is not accessible perturbatively due to the complex energy and the occurrence of exceptional lines in the high-field gap expression. For the two-dimensional toric code in a non-Hermitian staggered field we study the quantum robustness of the topologically ordered phase by the gap closing of the low-field gap. We find that the well-known second-order quantum phase transition of the toric code in a uniform field extends into a large portion of the non-Hermitian parameter space. However, the series expansions become unreliable for a dominant anti-Hermitian field. Interestingly, the analysis of the high-field gap reveals the potential presence of an intermediate region.
	\end{abstract}
	
	\maketitle
	
	%Introduction
	%%%%%%%%%%%%%%%%%%%%%%%%%%%%%%%%%%%%%%%%%%%%%%%%%%%%%%%%%%%%%%%%%%%%%%%%%%%%%%%%%%%%%%%%%%%%
	\section{Introduction}
	
	Non-Hermitian Hamiltonians with $PT$-symmetry represent an interesting extension of conventional quantum mechanics \cite{Bender_2007,Bender1998}, since such systems can display spontaneous $PT$-symmetry breaking from a purely real to a complex energy spectrum. Non-Hermitian operators typically arise when gain and loss terms are appropriately adjusted in open physical systems \cite{El-Ganainy_2018}. As a consequence, many experimental platforms are explored for non-Hermitian quantum dynamics like optical waveguides, electronics, microwaves, acoustics and single-spin systems \cite{Yuto_2020}. On the theoretical side, many investigations have focused either on single-particle or non-interacting non-Hermitian quantum systems, e.g., fermionic or bosonic band insulators with topologically non-trivial band structures \cite{Bergholtz_2021} while non-Hermitian interacting quantum many-body Hamiltonians are far less explored and understood.
	
	One class of non-Hermitian interacting quantum many-body systems are solvable one-dimensional quantum spin models \cite{Wang_2020,yang2020anomalous,Korff_2008,Castro_Alvaredo_2009,Albertini_1996,Liu_2021,Zhang_2013} like the Ising chain in a non-Hermitian staggered field \cite{Li2014}, which can be mapped in many cases to free fermions. However, higher-dimensional generalizations have not been treated to the best of our knowledge. Another class of quantum many-body systems are systems displaying intrinsic topological order which have a long-range entangled ground state and exotic anyonic excitations with non-trivial fractional statistics \cite{Leinaas_1977,Wilczek_1982}. One important question is the robustness of intrinsic topological order in non-Hermitian quantum systems. Here an attractive starting point are exactly solvable stabilizer codes like the toric code \cite{Kitaev2003}, which has been extended to non-Hermitian operators with $PT$-symmetry recently \cite{Guo_2020,Guo_2020b, Shackleton_2020}. It has been found that the correspondence between bulk quasi-particles and topologically protected degenerate ground states breaks down. Further, a continuous quantum phase transition without gap closing was explored that occurs in non-Hermitian topological orders \cite{Matsumoto_2020}.
	
Technically, one important tool to tackle quantum properties of interacting quantum spin systems are high-order series expansions \cite{Oitmaa_2006}, which have been used, among many other systems, to extract quantum critical properties in transverse-field Ising models \cite{He_1990,Weihong_1994,Powalski_2013,Coester2016,Roechner_2016} as well as for topological phase transitions in the toric code in the presence of external fields \cite{Vidal_2009,Dusuel2010,Dusuel_2011,Schmidt_2013}. However, these techniques have not been applied to non-Hermitian quantum spin models which is exactly the main purpose of this article. We extend high-order series expansions to investigate quantum critical properties of non-Hermitian quantum spin models. In particular, we calculate high-order series of energy gaps about different limits using the method of perturbative continuous unitary transformations \cite{Knetter2000,Knetter2003,Coester2016} or Takahashi's perturbation theory \cite{Takahashi1977} and apply extrapolations to extract critical points and associated critical exponents. In one dimension, we study the exactly solvable Ising chain in a non-Hermitian staggered field \cite{Li2014} which allows a direct comparison between the analytic solution and the findings from high-order series expansions. In two dimensions, we replace the Ising interaction by the toric code \cite{Kitaev2003} in order to understand the breakdown of the topological phase by the non-Hermitian staggered field.
	
	The article is structured as follows. In Sec.~\ref{sec:Models} we introduce the one-dimensional Ising chain and the two-dimensional toric code in a $PT$-symmetric non-Hermitian staggered field and we describe duality transformations for both systems to non-Hermitian staggered Ising interactions in a conventional transverse field. Technical aspects of the high-order low- and high-field series expansions and their extrapolations are given in Sec.~\ref{sec::se}. All results for the one-dimensional Ising chain in a non-Hermitian staggered field are contained in Sec.~\ref{sec::ising}. This includes the analytic solution as well as the comparison to series expansion results. In Sec.~\ref{sec::tc} we discuss the breakdown of the topological phase due to a non-Hermitian staggered Ising interaction. The article is concluded in Sec.~\ref{sec:conclusions}.
	 
	%Models
	%%%%%%%%%%%%%%%%%%%%%%%%%%%%%%%%%%%%%%%%%%%%%%%%%%%%%%%%%%%%%%%%%%%%%%%%%%%%%%%%%%%%%%%%%%%%
	\section{Non-Hermitian models} \label{sec:Models}
	
%	%Figure 1 - Ising chain
%	%%%%%%%%%%%%%%%%%%%%%%%%%%%%%%%%%%%%%%%%%%%%%%%%%%%%%%%%%%%%%%%%%%%%%%%%%%%%%%%%%%%%%%%%%%%%
%	\begin{figure}[t]
%		\centering
%		\def\svgwidth{\columnwidth}
%		\input{./figures/IsingChain.pdf_tex}
%		\caption{Ising chain in a non-Hermitian staggered transverse magnetic field. The sites represent the spin-{1/2} particles that are coupled by an Ising interaction of strength $J$. The colors encode the imaginary part of the magnetic prefactor that is given by $g = \eta + \imag \xi$ for empty red sites, belonging to $\circ$, and $g^* = \eta - \imag \xi$ for filled green sites, belonging to $\bullet$.}
%		\label{fig:ic}
%	\end{figure}
%	%%%%%%%%%%%%%%%%%%%%%%%%%%%%%%%%%%%%%%%%%%%%%%%%%%%%%%%%%%%%%%%%%%%%%%%%%%%%%%%%%%%%%%%%%%%%
%	
%	%Figure 2 - Toric code
%	%%%%%%%%%%%%%%%%%%%%%%%%%%%%%%%%%%%%%%%%%%%%%%%%%%%%%%%%%%%%%%%%%%%%%%%%%%%%%%%%%%%%%%%%%%%%
%	\begin{figure}[t]
%		\centering
%		\def\svgwidth{\columnwidth}
%		\input{./figures/ToricCode.pdf_tex}
%		\caption{Toric code in a non-Hermitian staggered parallel magnetic field. The sites represent the spin-{1/2} particles that are coupled by the toric code operators. The star operators $A_s$ are depicted in yellow, the plaquette operators $B_p$ in blue. The colors of the sites encode the imaginary part of the magnetic prefactor that is given by $g = \eta + \imag \xi$ for empty red sites, belonging to $\circ$, and $g^* = \eta - \imag \xi$ for filled green sites, belonging to $\bullet$.}
%		\label{fig:tc}
%	\end{figure}
%	%%%%%%%%%%%%%%%%%%%%%%%%%%%%%%%%%%%%%%%%%%%%%%%%%%%%%%%%%%%%%%%%%%%%%%%%%%%%%%%%%%%%%%%%%%%%
	
	We consider models with $N$ spin-$1/2$ particles on lattices that are coupled by some interaction Hamiltonian $\HInt$ and subject to a non-Hermitian staggered magnetic field. The prefactor of the magnetic field is either given by $g \equiv \eta + \imag \xi$ for spins on sublattice $\circ$ or by $g^* \equiv \eta - \imag \xi$ for spins on sublattice $\bullet$. The full Hamiltonian reads
	\begin{align}
		\mathcal{H} = \HInt - g \sum_{j \in \circ} \sigma^z_j - g^* \sum_{j \in \bullet} \sigma^z_j, \label{eq:Full_Hamiltonian}
	\end{align}
	where $\sigma^\alpha_j$ represents the Pauli matrix with flavor $\alpha$ on site $j$. The imaginary parts $\pm \xi$ of the magnetic field prefactors correspond to the anti-Hermitian part of the Hamiltonian. The assignment of spins to $\circ$ or $\bullet$ is done such that the full Hamiltonian is $\PT$-symmetric.
	
	For $\HInt = 0$, the so-called high-field limit, the model \eqref{eq:Full_Hamiltonian} is exactly solvable. The state with the lowest real eigenvalue $- |\eta| N$ is $\ket{\Uparrow}$. We will refer to this state as the ground state in the high-field limit. If additionally $\eta = 0$, the ground-state energy is infinitely degenerate in the thermodynamic limit; we will not treat this case with series expansions about the high-field limit. There are two types of elementary spin-flip excitations, depending on whether the flipped spin is on sublattice $\circ$ or $\bullet$. Their eigenvalues are given by $- |\eta| (N-2) \pm 2 \imag \xi$ and are non-real for $\xi \neq 0$.
	
	%Figure - Models
	%%%%%%%%%%%%%%%%%%%%%%%%%%%%%%%%%%%%%%%%%%%%%%%%%%%%%%%%%%%%%%%%%%%%%%%%%%%%%%%%%%%%%%%%%%%%
	\begin{figure}[t]
		\centering
		\def\svgwidth{\columnwidth}
		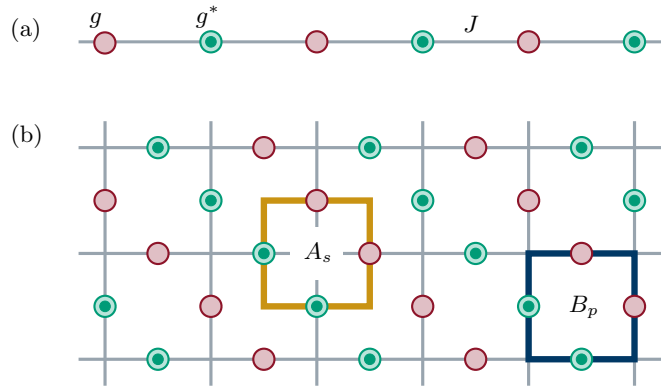
		\caption{Ising chain (a) and toric code (b) in a staggered magnetic field. The sites represent the spin-{1/2} particles that are coupled by an Ising interaction (a) or by the toric code operators (b). The colors of the sites encode the imaginary part of the magnetic prefactor that is given by $g = \eta + \imag \xi$ for empty red sites, belonging to $\circ$, and $g^* = \eta - \imag \xi$ for filled green sites, belonging to $\bullet$. The toric code star operators $A_s$ are depicted in yellow, the plaquette operators $B_p$ in blue.}
		\label{fig:models}
	\end{figure}
	%%%%%%%%%%%%%%%%%%%%%%%%%%%%%%%%%%%%%%%%%%%%%%%%%%%%%%%%%%%%%%%%%%%%%%%%%%%%%%%%%%%%%%%%%%%%
		
	We study two interaction Hamiltonians $\HInt$: the one-dimensional Ising chain and the 2D toric code \cite{Kitaev2003} on a square lattice. Both models are illustrated in Fig.~\ref{fig:models}. The full Hamiltonian of the Ising chain in a staggered transverse magnetic field reads
	\begin{align}
		\ic = - J \sum_{j=1}^N \sigma^x_j \sigma^x_{j+1} - g \sum_{j \in \circ} \sigma^z_j - g^* \sum_{j \in \bullet} \sigma^z_j, \label{eq:ic}
	\end{align}
	where we set $\sigma^x_{N+1} = \sigma^x_1$ and thus couple the chain periodically. The Ising terms $\sigma^x_j \sigma^x_{j+1}$ have eigenvalues $\pm 1$ depending on whether the respective spins are aligned or anti-aligned. These eigenvalues are not conserved as they do not commute with the Hamiltonian \eqref{eq:ic}. The Hermitian version of this model is exactly solvable \cite{Lieb1961, Pfeuty1970} using a Jordan-Wigner transformation \cite{Jordan1928} and a Bogoliubov transformation in Fourier space. Li et al.~\cite{Li2014} found a solution of the non-Hermitian generalization that they obtain via a Jordan-Wigner transformation and composite operators. In Sec.~\ref{sec:ic-exact_solution} we apply an alternative approach using a generalized Bogoliubov transformation in momentum space to derive the exact solution of the Ising chain in a non-Hermitian staggered magnetic field confirming their findings.
	
	The Hamiltonian of the toric code in a staggered parallel magnetic field reads
	\begin{align}
		\tc = - J \sum_s A_s - J \sum_p B_p - g \sum_{j \in \circ} \sigma^z_j - g^* \sum_{j \in \bullet} \sigma^z_j. \label{eq:tc}
	\end{align}
	Every star operator $A_s\equiv\prod_{i\in s}\sigma^x_i$ commutes with every plaquette operator $B_p\equiv\prod_{i\in p}\sigma^z_i$ because they always share an even number of spins. Thus all of them commute with the interaction Hamiltonian. The plaquette operators commute with the magnetic field terms as well and therefore also with the full Hamiltonian \eqref{eq:tc}. The star and plaquette operators have eigenvalues $\pm 1$ whereof the eigenvalues of the plaquette operators are conserved due to the afore mentioned commutation relations. That is why the Hamiltonian decomposes into blocks for the different combinations of plaquette eigenvalues. 
	
	The relevant low-energy physics takes place in the sector without plaquette excitations, i.e., the block where all plaquette operators have eigenvalue $\sgn{J}$. In this sector, the Hamiltonian simplifies to
	\begin{align}
		\tcg =  - |J| \frac{N}{2} - J \sum_s A_s - g \sum_{j \in \circ} \sigma^z_j - g^* \sum_{j \in \bullet} \sigma^z_j. \label{eq:tcg}
	\end{align}
	
	For both systems \eqref{eq:ic} and \eqref{eq:tcg} it suffices to study a restricted parameter space. We may without loss of generality choose $\xi \geq 0$, because any sign in front of $\xi$ can be absorbed into a reassignment of spins to $\circ$ and $\bullet$. We may also choose $J, \eta \geq 0$ because any sign in front of $J$ or in front of both, $\eta$ and $\xi$, can be absorbed into a rotation in spin space of some spins around the $z$-axis or all spins around the $x$-axis, respectively.
	
	Interestingly, both models are dual to the transverse field Ising model with non-Hermitian staggered Ising interaction. The dual model of the Ising chain \eqref{eq:ic} lives on a chain with effective sites centered on the links of the original model. Similarly, for the toric code \eqref{eq:tcg} in the subspace with all plaquette operators having eigenvalues $+1$, the dual lattice is a square lattice build by the centers of star operators. The effective sites are labeled by $\mu$ and $\nu$. Because the interaction Hamiltonian contains in both cases only operators with eigenvalues $\pm 1$, this mapping uses Pauli matrices $\tau^z$ for pseudo-spins 1/2 that are defined to have the same eigenvalues so that $\HInt$ is mapped to an effective magnetic field in $z$-direction. Further, each local operator of the magnetic field term changes the eigenvalues of two adjacent pseudo-spins on sites $\mu$ and $\nu$ which therefore gives an effective Ising interaction $\tau^x_\mu \tau^x_\nu$. The resulting dual Hamiltonian reads then in both cases
	\begin{align}
		\mathcal{H} = \bar{E}_0 - J \sum_\mu \tau^z_\mu - g \sum_{\langle \mu, \nu \rangle \in \circ} \tau^x_\mu \tau^x_\nu - g^* \sum_{\langle \mu, \nu \rangle \in \bullet} \tau^x_\mu \tau^x_\nu, \label{eq:TFIM}
	\end{align}
	where $\langle \mu, \nu \rangle$ represents bonds of the dual lattice. The assignment of bonds to the former sublattices $\circ$ and $\bullet$ represent the colors of the bonds. Note that the explicit value of the constant $\bar{E}_0$ differs for the two cases, but plays no role physically.
	
	%Series expansions
	%%%%%%%%%%%%%%%%%%%%%%%%%%%%%%%%%%%%%%%%%%%%%%%%%%%%%%%%%%%%%%%%%%%%%%%%%%%%%%%%%%%%%%%%%%%%
	\section{Series expansions}\label{sec::se}
	In order to explore the ground-state phase diagrams, we perform high-order series expansions about the low-field limit, where $\eta = \xi = 0$, and about the high-field limit, where $J = 0$. We will test the series expansion by comparing to the exactly solvable Ising chain in a non-Hermitian staggered transverse magnetic field given in Sec.~\ref{sec:ic-exact_solution}. The goal of these methods is to transform the Hamiltonians into effective quasi-particle (qp) conserving ones. Being quasi-particle conserving, the effective Hamiltonians decompose into decoupled blocks for different quasi-particle numbers, which enables us to obtain access to the relevant low-energy physics, i.e., the ground-state energy via the zero qp block and the one-particle gap via the one qp block.
	
	%pCUT
	%%%%%%%%%%%%%%%%%%%%%%%%%%%%%%%%%%%%%%%%%%%%%%%%%%%%%%%%%%%%%%%%%%%%%%%%%%%%%%%%%%%%%%%%%%%%
	\subsection{Low-field expansions} \label{sec:Low_Field}
	In the following we study the low-field (lf) limit $g\ll J$ perturbatively. We introduce hardcore-bosonic operators 
	\begin{align}
		b^{\phantom{\dagger}}_\mu &= (\tau^x_\mu + \imag \tau^y_\mu)/2, & b^\dagger_\mu &= (\tau^x_\mu - \imag \tau^y_\mu)/2
	\end{align}
	to express Eq.~\eqref{eq:TFIM} exactly as follows 
	\begin{align}
		\mathcal{H} &= - J N + 2 J \sum_{\mu} b^\dagger_\mu b^{\phantom{\dagger}}_\mu - g \sum_{\langle \mu, \nu \rangle \in \circ} \left( b^\dagger_\mu + b^{\phantom{\dagger}}_\mu \right) \left( b^\dagger_\nu + b^{\phantom{\dagger}}_\nu \right) \nonumber\\
		&\hphantom{{}=} - g^* \sum_{\langle \mu, \nu \rangle \in \bullet} \left( b^\dagger_\mu + b^{\phantom{\dagger}}_\mu \right) \left( b^\dagger_\nu + b^{\phantom{\dagger}}_\nu \right),
	\end{align}
	where $\mathcal{Q} \equiv \sum_{\mu} b^\dagger_\mu b^{\phantom{\dagger}}_\mu$ is a counting operator reflecting the equidistant spectrum of the unperturbed Hamiltonian $g=0$. The unperturbed ground state $\ket{0}_\lf$ is then defined by $\mathcal{Q} \ket{0}_\lf = 0$. Rescaling the Hamiltonian by $2J$ yields
	\begin{align}
		\frac{\mathcal{H}}{2J} = - \frac{N}{2} + \mathcal{Q} - \frac{g}{2J} \sum_{n \in \{-2, 0, 2\}} T_n^\circ - \frac{g^*}{2J} \sum_{n \in \{-2, 0, 2\}} T_n^\bullet, \label{eq:TFIM-pCUT}
	\end{align}
	where $[\mathcal{Q}, T_n^{\circ/\bullet}] = n T_n^{\circ/\bullet}$, i.e. $T_n^{\circ/\bullet}$ creates $n$ elementary excitations. Every operator used here is Hermitian, the anti-Hermitian parts arising only from the imaginary part $\pm \xi$ of the magnetic field prefactors. This aspect is essential for the straightforward generalization of the method of perturbative continuous unitary transformations (pCUTs) \cite{Knetter2000, Knetter2003} to non-Hermitian Hamiltonians \eqref{eq:Full_Hamiltonian}.
	
	For Hermitian Hamiltonians, the method of continuous unitary transformations \cite{Wegner1994} works by introducing a unitary transformation $U(\ell)$ that depends on a parameter $\ell$ flowing continuously from $0$ to $\infty$. The so called flow Hamiltonian $\mathcal{H}(\ell) = U^{-1}(\ell) \mathcal{H} U(\ell)$ is also $\ell$-dependent and becomes the desired effective Hamiltonian as $\ell$ approaches $\infty$. The unitary transformation is not explicitly needed. It suffices to introduce its anti-Hermitian infinitesimal generator $\eta(\ell)$. The effective Hamiltonian can be calculated using the flow equation
	\begin{align}
		\frac{\textrm{d} \mathcal{H}(\ell)}{\textrm{d} \ell} = [\eta(\ell), \mathcal{H}(\ell)] \label{eq:pCUT}
	\end{align}
	and taking the limit $\ell$ to $\infty$. If the Hermitian Hamiltonian consists of a counting operator $\mathcal{Q}$ representing the unperturbed Hamiltonian and a finite number of $T_n$ operators with $[\mathcal{Q},T_n]=nT_n$ connected to one or more perturbation parameters, the pCUT method is applicable. For models with one perturbation parameter $\lambda$, the ansätze for the infinitesimal generator and the flow Hamiltonian in the pCUT method are
	\begin{widetext}
	\begin{align}
		\eta(\ell) &= \sum_{k=1}^{\infty} \lambda^k \sum_{(m_1, \ldots, m_k)} \sgn{m_1 + \cdots + m_k} F(\ell; (m_1, \ldots, m_k)) T_{m_1} \cdots T_{m_k}, \label{eq:pCUT-eta}\\
		\mathcal{H}(\ell) &= \mathcal{Q} + \sum_{k=1}^{\infty} \lambda^k \sum_{(m_1, \ldots, m_k)} F(\ell; (m_1, \ldots, m_k)) T_{m_1} \cdots T_{m_k}, \label{eq:pCUT-H}
	\end{align}
	\end{widetext}
	where $F$ are real-valued coefficient functions. The generalization to multiple perturbation parameters is straightforward. Inserting Eqs.~\eqref{eq:pCUT-eta} and \eqref{eq:pCUT-H} into the flow equation \eqref{eq:pCUT}, the functions $F$ can be calculated recursively order by order. This can be done computer-aided up to high orders resulting in an effective Hamiltonian $\mathcal{H}_\textrm{eff}$ for $\ell\rightarrow\infty$ with rational coefficients which is blockdiagonal in the qp number, i.e., $[\mathcal{H}_\textrm{eff}, \mathcal{Q}] = 0$ holds. One advantage of this method is that it is not model specific, i.e., the solution for the coefficient functions and therefore for the effective Hamiltonian does only depend on the number and type of the $T_n$ operators. The specific underlying model is only needed for explicitly calculating quantities like the ground-state energy or the gap, which corresponds to normal-ordering the effective pCUT Hamiltonian $\mathcal{H}_\textrm{eff}$.
	
	If a Hamiltonian is not Hermitian but has a complex perturbation parameter instead, the flow equation \eqref{eq:pCUT} still holds. Therefore the ansätze \eqref{eq:pCUT-eta} and \eqref{eq:pCUT-H} do not have to be altered and the calculated coefficient functions remain unchanged. Note that in this case the infinitesimal generator $\eta(\ell)$ is not anti-Hermitian and the transformation $U(\ell)$ not unitary but a similarity transformation. Non-perturbative similarity transformations using the quasi-particle generator were already applied successfully for non-Hermitian Hamiltonians \cite{Powalski2015,Powalski2018} to describe magnetic excitations of the long-range ordered Heisenberg model on the square lattice using the Dyson-Maleev transformation. They are referred to as CST for continuous similarity transformation. For our systems, the Hamiltonian can be written in both cases as Eq.~\eqref{eq:TFIM-pCUT} with the two perturbation parameters $-g/(2J)$ and $-g^*/(2J)$. In order to obtain the respective ground-state energy and one-particle gap, we have calculated the zero and one qp block of $\mathcal{H}_\textrm{eff}^\lf$ for both models.
	
	The unperturbed ground-state energy $-JN$ is non-degenerate in both cases, thus the zero qp block is a number equal to the perturbed ground-state energy. It is given by
	\begin{align}
		E_0^\lf = \tensor*[_\lf]{\bra{0}}{} \mathcal{H}_\textrm{eff}^\lf \ket{0}_\lf.
	\end{align}
	Its specific value has to be calculated for both models separately. The results are given in the appendix~\ref{sec:Appendix-Low_Field} in Eq.~\eqref{eq:ic-Low_Field-Ground_State_Energy} [Eq.~\eqref{eq:tcg-Low_Field-Ground_State_Energy}] up to \nth{12} order for the 1D [2D] case. For the 2D case we have applied a full graph decomposition.
	
	The one qp block consists of the elements
	\begin{align}
		\bra{0} b^{\phantom{\dagger}}_\mu \mathcal{H}_\textrm{eff}^\lf\, b^\dagger_\nu \ket{0} = E_0^\lf \delta_{\mu\nu} + t_{\mu\nu},
	\end{align}
	where $t_{\mu\nu}$ denotes the hopping amplitude from $\nu$ to $\mu$. After a Fourier transform	this block further decomposes into decoupled $2 \times 2$ blocks for different momenta $k$. The two bands of the dispersion relation are obtained by diagonalizing these blocks separately. The gap closes at $k = 0$ for both systems, so we only have to diagonalize the corresponding block and pick the lower eigenvalue. The respective series are given in appendix~\ref{sec:Appendix-Low_Field} in Eq.~\eqref{eq:ic-Low_Field-Gap} [Eq.~\eqref{eq:tcg-Low_Field-Gap}] up to \nth{10} order for the 1D [2D] case. The 2D results were again obtained using a full graph decomposition.
	
	%Takahashi
	%%%%%%%%%%%%%%%%%%%%%%%%%%%%%%%%%%%%%%%%%%%%%%%%%%%%%%%%%%%%%%%%%%%%%%%%%%%%%%%%%%%%%%%%%%%%
	\subsection{High-field expansions} \label{sec:High_Field}
	In the following we study the high-field (hf) limit $J\ll g$. The unperturbed case $J = 0$ corresponds to $\HInt = 0$. This limit was already brought up in Sec.~\ref{sec:Models}. Similar to before, we introduce hardcore-bosonic operators
	\begin{align}
		b^{\phantom{\dagger}}_j &= (\sigma^x_j + \imag \sigma^y_j)/2 & b^\dagger_j &= (\sigma^x_j - \imag \sigma^y_j)/2
	\end{align}
	and express the full Hamiltonians \eqref{eq:ic} and \eqref{eq:tcg} as
	\begin{align}
		\begin{split}
			\ic &= - \eta N + 2 g \sum_{j \in \circ} b^\dagger_j b^{\phantom{\dagger}}_j + 2 g^* \sum_{j \in \bullet} b^\dagger_j b^{\phantom{\dagger}}_j\\
			&\hphantom{{}=} - J \sum_{\langle i, j \rangle} \left( b^\dagger_i + b^{\phantom{\dagger}}_i \right) \left( b^\dagger_j + b^{\phantom{\dagger}}_j \right), \label{eq:ic-High_Field}
		\end{split}\\
		\begin{split}
			\tcg &= - \eta N + 2 g \sum_{j \in \circ} b^\dagger_j b^{\phantom{\dagger}}_j + 2 g^* \sum_{j \in \bullet} b^\dagger_j b^{\phantom{\dagger}}_j\\
			&\hphantom{{}=} - J \frac{N}{2} - J \sum_s \prod_{j \in s} \left( b^\dagger_j + b^{\phantom{\dagger}}_j \right). \label{eq:tcg-High_Field}
		\end{split}
	\end{align}
	When \mbox{$\eta \neq 0$}, the unperturbed ground state is unique in both cases, defined by \mbox{$\sum_j b^\dagger_j b^{\phantom{\dagger}}_j \ket{0}_\hf = 0$} and corresponds to the state $\ket{\Uparrow}$ introduced in Sec.~\ref{sec:Models}. It is not possible to rescale these Hamiltonians such that the unperturbed parts are expressed as counting operators. This implies that we cannot use the same method as in the low-field limit. Instead, we use Takahashi's perturbation theory \cite{Takahashi1977} that is based on \cite{Kato1949}.
	
	This method is suited for Hamiltonians of the form \mbox{$\mathcal{H} = H_0 + \lambda V$} with perturbation parameter $\lambda$. Assume that the unperturbed part has an eigenvalue $E$ whose perturbative corrections we are interested in. Let $P_0$ be a projector onto the space spanned by the unperturbed eigenvectors. We can then construct a projector onto the space spanned by the perturbed eigenvectors corresponding to the perturbed eigenvalue by
	\begin{align}
		P = P_0 - \sum_{k = 1}^{\infty} \lambda^k \sum_{k_1 + \cdots + k_{k+1} = k, k_i \geq 0} S^{k_1} V \cdots V S^{k_{k+1}}
	\end{align}
	with $S = (1 - P_0)/(E - H_0)$ being a projector onto the complement of the space spanned by the unperturbed eigenvectors and $S^0 = -P_0$. The transformation that generates the effective Hamiltonian is obtained by expanding $\Gamma = P P_0 (P_0 P P_0)^{-1/2}$ in powers of $\lambda$. 
	
	The ground-state energy and gap for both models are obtained analogously to Sec.~\ref{sec:Low_Field}. Because of $\eta \neq 0$ the unperturbed ground-state energy is non-degenerate and the perturbed ground-state energy can be calculated by
	\begin{align}
		E_0^\hf = \tensor[_\hf]{\bra{0}}{} \mathcal{H}_\textrm{eff}^\hf \ket{0}_\hf.
	\end{align}
	The series are given in appendix~\ref{sec:Appendix-High_Field} in Eq.~\eqref{eq:ic-High_Field-Ground_State_Energy} [Eq.~\eqref{eq:tcg-High_Field-Ground_State_Energy}] up to \nth{10} [\nth{6}] order for the 1D [2D] case.
	
	In the 1D case the one qp block consists of the elements
	\begin{align}
		\tensor[_\hf]{\bra{0}}{} b^{\phantom{\dagger}}_i \mathcal{H}_\textrm{eff}^\hf\, b^\dagger_j \ket{0}_\hf = E_0^\hf \delta_{ij} + t_{ij},
	\end{align}
	where $t_{ij}$ denotes the hopping amplitude from $j$ to $i$. For the 2D case the relevant excitation is composed of four spin flips on a star as all excitations with less spin flips violate the constraint that all eigenvalues of plaquette operators are $+1$. The respective qp block consists of the elements
	\begin{align}
		\tensor[_\hf]{\bra{0}}{} \left( \prod_{i \in s} b^{\phantom{\dagger}}_i \right) \mathcal{H}_\textrm{eff}^\hf\, \left( \prod_{j \in s'} b^\dagger_j \right) \ket{0}_\hf = E_0^\hf \delta_{ss'} + t_{ss'},
	\end{align}
	where $t_{ss'}$ denotes the hopping amplitude of the composite entity from star $s'$ to $s$.
	After a Fourier transform this block further decomposes into decoupled $2 \times 2$ blocks for different momenta $k$. The two bands of the dispersion relation are obtained by diagonalizing these blocks separately.
	
	The energy of the elementary excitations with $k = 0$ in the 1D case is given in the appendix~\ref{sec:Appendix-High_Field} in Eq.~\eqref{eq:ic-High_Field-Gap} up to \nth{10} order. The gap for the 2D case is given up to \nth{4} order in Eq.~\eqref{eq:tcg-High_Field-Gap}. For the special case $\xi = 0$ the gap is known up to even higher orders \cite{Dusuel2010} and the series up to order 10 is presented in Eq.~\eqref{eq:tcg-High_Field-Gap-Hermitian} for completeness. In the parameter region defined by $\eta = \xi$ we have further calculated the gap explicitly up to \nth{8} order in Eq.~\eqref{eq:tcg-High_Field-Gap-Diagonal}.
	
	%DLog Padés
	%%%%%%%%%%%%%%%%%%%%%%%%%%%%%%%%%%%%%%%%%%%%%%%%%%%%%%%%%%%%%%%%%%%%%%%%%%%%%%%%%%%%%%%%%%%%
	\subsection{DLog Padé approximants}
	In the vicinity of a second-order quantum critical point $\crit$, the gap typically vanishes as $\Delta \propto |\crit - x|^{z\nu}$ with the dynamical critical exponent $z$ and the correlation length critical exponent $\nu$. Similarly, the second derivative of the ground-state energy diverges according to a power law $\text{d}^2 e_0 / \text{d} x^2 \propto |\crit - x|^{-\alpha}$ for $\alpha\neq 0$ with the specific heat critical exponent $\alpha$.
	
	Whenever we have a function $F$ with such a power law behavior, we can use the logarithmic derivative in order to extract the value of the critical exponent $\vartheta$ (see \cite{Guttmann1989} for a general introduction of extrapolation techniques like Padé and DLog Padé approximation). Assume that close to the critical point $\crit$
	\begin{align}
		F(x) \approx |\crit - x|^{-\vartheta} A(x)
	\end{align}
	with $A$ being analytic at $\crit$. This enables us to approximate $A(x) \approx A(\crit) \left(1 + \mathcal{O} (\crit - x) \right)$ and to calculate the logarithmic derivative
	\begin{align}
		D(x) = \frac{\text{d}}{\text{d} x} \ln(F(x)) \approx \frac{\vartheta}{\crit - x} \left(1 + \mathcal{O} (\crit - x) \right)
	\end{align}
	close to the critical point. The critical exponent is then given by the residue of this derivative at the critical point.
	
	If the function $F$ is not known but only its Taylor expansion about $x = 0$ up to some order $k$, we use DLog Padé approximants to estimate the critical point and corresponding exponent. The DLog Padé approximant of $F$ of order $(L, M)$ with $L + M = k - 1$ is defined by
	\begin{align}
		P_\text{DLog}[L, M]_F(x) = P[L, M]_D(x),
	\end{align}
	where
	\begin{align}
		P[L, M]_D(x) = \frac{P_L(x)}{Q_M(x)} = \frac{p_0 + p_1 x + \cdots + p_L x^L}{1 + q_1 x + \cdots + q_M x^M}
	\end{align}
	is the Padé approximant of order $(L, M)$ of the logarithmic derivative. The coefficients are fixed by imposing that the Taylor series of the Padé approximant coincides with the one of the logarithmic derivative up to order $k - 1$. A DLog Padé approximant indicates the critical point to be the root of its denominator $Q_M(x)$ and the critical exponent to be its residue at the critical point.
	
	It is possible to obtain an approximation of the function $F$ by
	\begin{align}
		F(x) \approx F(0) \exp\left( \int_{0}^{x} \diff y\, P_\text{DLog}[L, M]_F(y) \right). \label{eq:Integrated-Pade}
	\end{align}

	If the critical exponent $\vartheta$ is known we can perform a biased DLog Padé approximation, i.e., we add the next highest order (the next even order in case the series contains only even orders) to the Taylor series of $F$ and take the respective prefactor $a$ such that the critical exponent is the desired one. This prefactor can differ depending on the order $(L, M)$ of the approximant we are interested in.
	
	For our analysis, we calculate all possible DLog Padé approximants with $L, M \geq 2$ and group them into families defined by constant $L - M$. Approximants with removable root-pole pairs have to be excluded as they carry the same information as their lower-ordered relatives and thus convey a false sense of convergence. If the denominator of an approximant has additional roots close to the expected one, the approximant has to be discarded as defective as well, because the additional root may deform it. Here, we also have to take complex roots into account and specify that additional roots are considered to be close to the expected root $\crit$ if their distance in the complex plane is smaller than $\crit/2$ . Finally, we might only include families with at least two remaining members in our analysis, depending on the total number of available approximants.
	
	The critical exponents of the Hermitian limits of \eqref{eq:TFIM} are either known exactly for the 1D TFIM or estimated with high precision for the 2D TFIM, which then naturally translates to the dual toric code in a field. For the 1D chain, the critical exponents are $\nu = z= 1$ and the specific heat critical exponent is $\alpha = 0$, i.e., the second derivative of the ground-state energy diverges logarithmically and not according to a power law. The best estimates for the exponents of the 2D model are obtained via conformal bootstrap or Monte Carlo simulations; the results given are $z = 1$ \cite{Pfeuty1971, Kos2016}, $\nu = 0.629971(4)$ \cite{Kos2016}, and $\alpha = 0.110087(12)$ \cite{Kos2016}.
	
	%Ising chain - Results
	%%%%%%%%%%%%%%%%%%%%%%%%%%%%%%%%%%%%%%%%%%%%%%%%%%%%%%%%%%%%%%%%%%%%%%%%%%%%%%%%%%%%%%%%%%%%
	\section{Ising chain results}\label{sec::ising}
	
	%Ising chain - Exact solution
	%%%%%%%%%%%%%%%%%%%%%%%%%%%%%%%%%%%%%%%%%%%%%%%%%%%%%%%%%%%%%%%%%%%%%%%%%%%%%%%%%%%%%%%%%%%%
	\subsection{Exact solution} \label{sec:ic-exact_solution}
	The Ising chain in a Hermitian uniform transverse magnetic field is exactly solvable \cite{Pfeuty1970, Lieb1961} by using a Jordan-Wigner transformation \cite{Jordan1928} and a Bogoliubov transformation in Fourier space. The non-Hermitian generalization \eqref{eq:ic} of this model is solvable by using the same steps, adapted to a larger unit cell.
	
	Using $\sigma^\pm_j = (\sigma^x_j \pm \sigma^y_j)/2$, we define the fermionic operators of the Jordan-Wigner transformation by
	\begin{align}
		c^{\phantom{\dagger}}_j &:= \left(\prod_{l=1}^{j-1} \sigma^z_l\right) \sigma^+_j, & c^\dagger_j &:= \left(\prod_{l=1}^{j-1} \sigma^z_l\right) \sigma^-_j \label{eq:ic-Jordan_Wigner_Operators}
	\end{align}
	for $j \in \{1, \ldots, N\}$. Inverted, the Pauli operators are given by
	\begin{align}
		\sigma^z_j &= 1 - 2 \, c^\dagger_j c^{\phantom{\dagger}}_j,\\
		\sigma^x_j &= \prod_{l=1}^{j-1} \left( 1 - 2 \, c^\dagger_l c^{\phantom{\dagger}}_l \right) \left( c^\dagger_j + c^{\phantom{\dagger}}_j \right)
	\end{align}
	for $j \in \{1, \ldots, N\}$. The Hamiltonian expressed in terms of the fermionic operators \eqref{eq:ic-Jordan_Wigner_Operators} reads 
	\begin{align}
		\begin{split}
			\ic &= - J \sum_{j=1}^N \left( c^\dagger_j c^{\phantom{\dagger}}_{j+1} + c^\dagger_j c^\dagger_{j+1} + \textrm{h.c.} \right)\\
			&\hphantom{{}=} - N \eta + 2 \sum_{j=1}^N g^{\phantom{*}}_j c^\dagger_j c^{\phantom{\dagger}}_j,
		\end{split}
	\end{align}
	where we have neglected a boundary term that vanishes in the thermodynamic limit \cite{Lieb1961}. This term stems from transforming $\sigma^x_N \sigma^x_1$ and it arises from the fact that we impose $c_{N+1} = c_1$ which might differ by a sign from inserting $j=N+1$ into Eq.~\eqref{eq:ic-Jordan_Wigner_Operators}.
	
	We take the Fourier transform
	\begin{align}
		c^{\phantom{\dagger}}_{k, \circ} &= \sqrt{\frac{2}{N}} \sum_{j \in \circ} e^{-\imag k(j-1)/2} c^{\phantom{\dagger}}_j\\
		c^{\phantom{\dagger}}_{k, \bullet} &= \sqrt{\frac{2}{N}} \sum_{j \in \bullet} e^{-\imag kj/2} c^{\phantom{\dagger}}_j,
	\end{align}
	where $\circ$ contains all empty spins marked in red and $\bullet$ all filled spins marked in green in Fig.~\hyperref[fig:models]{\ref*{fig:models}(a)}. The Hamiltonian decomposes into a sum $\ic = \sum_k H_k$ with
	\begin{widetext}
	\begin{align}
		H_k &= - J \left( c^\dagger_{k, \circ} c^{\phantom{\dagger}}_{k, \bullet} + e^{\imag k} c^\dagger_{k, \bullet} c^{\phantom{\dagger}}_{k, \circ} + c^\dagger_{k, \circ} c^\dagger_{-k, \bullet} + e^{\imag k} c^\dagger_{k, \bullet} c^\dagger_{-k, \circ} + \textrm{h.c.} \right) - 2 \eta + 2 (\eta + \imag \xi) c^\dagger_{k, \circ} c^{\phantom{\dagger}}_{k, \circ} + 2 (\eta - \imag \xi) c^\dagger_{k, \bullet} c^{\phantom{\dagger}}_{k, \bullet}.
	\end{align}
	\end{widetext}
	
	The transformed Hamiltonian is now block-diagonal with blocks $h_k = H_k + H_{-k}$. These blocks are individually diagonalizable using a generalized Bogoliubov transformation. We define vectors of fermionic operators
	\begin{align}
		\mathbf{c}_k = \begin{pmatrix}
		c^{\phantom{\dagger}}_{k, \circ} & c^{\phantom{\dagger}}_{k, \bullet} & c^{\phantom{\dagger}}_{-k, \circ} & c^{\phantom{\dagger}}_{-k, \bullet}
		\end{pmatrix}
	\end{align}
	to write the blocks as
	\begin{align}
		h_k = \frac{1}{2} \begin{pmatrix}
		\mathbf{c}^\dagger_k & \mathbf{c}^{\phantom{\dagger}}_k
		\end{pmatrix} \begin{pmatrix}
		-\mathbf{A}_k & \mathbf{B}_k\\
		\mathbf{B}^\dagger_k & \mathbf{A}^\textrm{T}_k
		\end{pmatrix} \begin{pmatrix}
		\mathbf{c}^{\phantom{\dagger}}_k\\
		\mathbf{c}^\dagger_k
		\end{pmatrix},
	\end{align}
	where the matrices $\mathbf{A}_k$ and $\mathbf{B}_k$ are the components of the coefficient matrix. $\mathbf{A}_k$ is given by
	\begin{align}
		\begin{pmatrix}
		-2 (\eta + \imag \xi) & J (e^{\imag k}+1) & 0 & 0\\
		J (e^{-\imag k}+1) & -2 (\eta + \imag \xi) & 0 & 0\\
		0 & 0 & 2 (\eta + \imag \xi) & J (e^{-\imag k}+1)\\
		0 & 0 & J (e^{\imag k}+1) & 2 (\eta - \imag \xi)
		\end{pmatrix}
	\end{align}
	and $\mathbf{B}_k$ by
	\begin{align}
		\begin{pmatrix}
		0 & 0 & 0 & -J (e^{\imag k}-1)\\
		0 & 0 & J (e^{-\imag k}-1) & 0\\
		0 & -J (e^{-\imag k}-1) & 0\\
		J (e^{\imag k}-1) & 0 & 0 & 0
		\end{pmatrix}.
	\end{align}
	For a Hermitian problem, the matrices $\mathbf{A}_k$ would be Hermitian and the matrices $\mathbf{B}_k$ anti-symmetric. This would make the coefficient matrix Hermitian and therefore unitary diagonalizable. Our matrix is not Hermitian but normal and therefore still unitary diagonalizable. The corresponding transformation maps the fermionic operators $c$ onto other fermionic operators $\gamma$. The so diagonalized Hamiltonian takes the form
	\begin{align}
		\begin{split}
			\ic &= - \sum_k (\omega^+_k + \omega^-_k)\\
			&+ 2 \sum_k \left( \omega^+_k \gamma^\dagger_{k, +} \gamma^{\phantom{\dagger}}_{k, +} + \omega^-_k \gamma^{\dagger}_{k, -} \gamma^{\phantom{\dagger}}_{k, -} \right)
		\end{split}
	\end{align}
	with the two bands
	\begin{widetext}
	\begin{align}
		\omega^\pm_k &= \sqrt{J^2 + \eta^2 - \xi^2 \pm \sqrt{2 J^2 (\eta^2 - \xi^2) - 4 \eta^2 \xi^2 + 2 J^2 (\eta^2 + \xi^2) \cos(k)}} \label{eq:ic-dispersion-cartesian}\\
		&= J \sqrt{1 + r^2 \cos(2 \varphi) \pm r \sqrt{2 \cos(2 \varphi) - r^2 \sin^2(2 \varphi) + 2 \cos(k)}}, \label{eq:ic-dispersion-radial}
	\end{align}
	\end{widetext}
	where Eq.~\eqref{eq:ic-dispersion-radial} is obtained by inserting $\eta = J r \cos(\varphi)$ and $\xi = J r \sin(\varphi)$ into Eq.~\eqref{eq:ic-dispersion-cartesian}. These bands can also be obtained through linear combination of results by Li et al. \cite{Li2014}.
	
	%Ising chain - Phase diagram
	%%%%%%%%%%%%%%%%%%%%%%%%%%%%%%%%%%%%%%%%%%%%%%%%%%%%%%%%%%%%%%%%%%%%%%%%%%%%%%%%%%%%%%%%%%%%
	\subsection{Phase diagram}
	
	%Figure - Ising chain - Phase Diagram
	%%%%%%%%%%%%%%%%%%%%%%%%%%%%%%%%%%%%%%%%%%%%%%%%%%%%%%%%%%%%%%%%%%%%%%%%%%%%%%%%%%%%%%%%%%%%
	\begin{figure}[t]
		\centering
		\resizebox{\columnwidth}{!}{\input{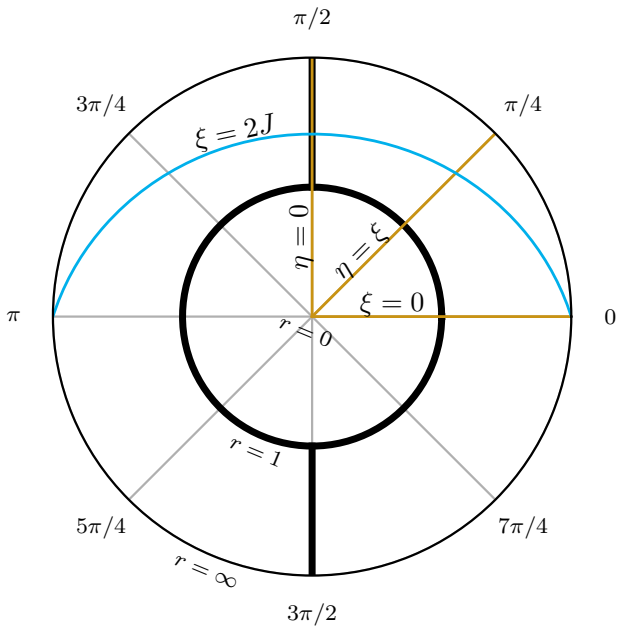}}
		\caption{Phase diagram of the non-Hermitian Ising chain in a staggered transverse magnetic field \eqref{eq:ic} as a function of $\arctan(r)$ and $\varphi$ in polar coordinates. The ground-state energy diverges logarithmically at the boundaries indicated by thick black lines. The yellow radial lines correspond to paths with $\xi=0$, $\eta=\xi$, and $\eta=0$ displayed in Fig.~\ref{fig:ic-gap-radial} while the blue line refers to the path with $\xi = 2J$ displayed in Fig.~\ref{fig:ic-gap-line}.}
		\label{fig:ic-phase_diagram}
	\end{figure}
	%%%%%%%%%%%%%%%%%%%%%%%%%%%%%%%%%%%%%%%%%%%%%%%%%%%%%%%%%%%%%%%%%%%%%%%%%%%%%%%%%%%%%%%%%%%%
	
	The analytic solution to our model enables us to investigate the phase transitions, as well as the points where energies become complex. There are energies that are real in the whole parameter space and energies that are real only in certain regions. The only region with real spectrum is the Hermitian axis given by $\xi=0$.
		
	The ground-state energy per spin is given by
	\begin{align}
		e_0 = - \frac{1}{4\pi} \int\limits_0^{2\pi} (\omega^+_k + \omega^-_k) \diff k
	\end{align}
	in the thermodynamic limit. This energy is real, because the imaginary parts of the bands $\omega^+_k$ and $\omega^-_k$ cancel each other. Li et al.~\cite{Li2014} argue that the Laplacian of the ground-state energy diverges logarithmically at \mbox{$r = 1$} and at $\eta = 0$ for $\xi \geq J$. This indicates second-order phase transitions at these boundaries, as depicted in Fig.~\ref{fig:ic-phase_diagram}.

	%Figure - Ising chain - Gap - Radial
	%%%%%%%%%%%%%%%%%%%%%%%%%%%%%%%%%%%%%%%%%%%%%%%%%%%%%%%%%%%%%%%%%%%%%%%%%%%%%%%%%%%%%%%%%%%%
	\begin{figure}[t]
		\centering
		\resizebox{\columnwidth}{!}{\input{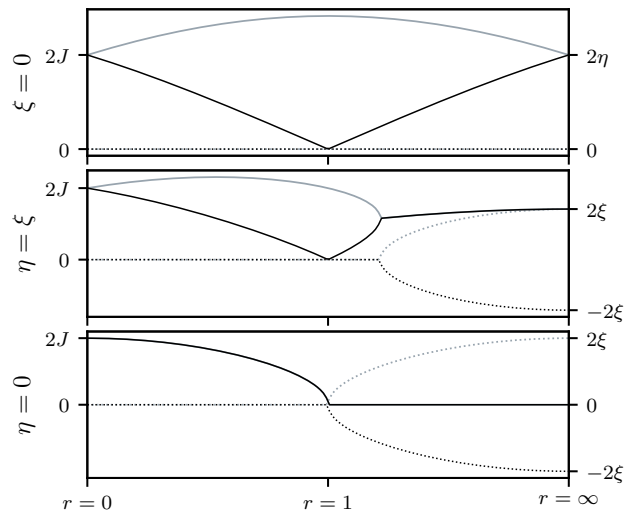}}
		\caption{Energies above the ground-state energy of elementary excitations with momentum $k=0$ of the non-Hermitian Ising chain in a staggered transverse magnetic field as a function of $\arctan(r)$ between $0$ and $\pi/2$. The three plots correspond to the three yellow radial paths in Fig.~\ref{fig:ic-phase_diagram} with $\xi=0$ (upper panel), $\eta = \xi$ (middle panel), and $\eta=0$ (lower panel). Real parts are depicted using solid lines, imaginary parts using dotted lines. The gap $\Delta = 2 \omega^-_0$ is depicted in black, the other elementary excitation with the same momentum $2 \omega^+_0$ in gray. The gap closes at $r=1$, indicating a second-order phase transition. The energies are real if and only if $\xi \leq J$.}
		\label{fig:ic-gap-radial}
	\end{figure}
	%%%%%%%%%%%%%%%%%%%%%%%%%%%%%%%%%%%%%%%%%%%%%%%%%%%%%%%%%%%%%%%%%%%%%%%%%%%%%%%%%%%%%%%%%%%%
	
	%Figure - Ising chain - Gap - Line
	%%%%%%%%%%%%%%%%%%%%%%%%%%%%%%%%%%%%%%%%%%%%%%%%%%%%%%%%%%%%%%%%%%%%%%%%%%%%%%%%%%%%%%%%%%%%
	\begin{figure}[t]
		\centering
		\resizebox{\columnwidth}{!}{\input{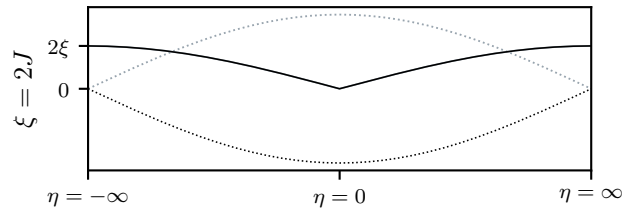}}
		\caption{Energies above the ground-state energy of elementary excitations with momentum $k=0$ of the non-Hermitian Ising chain in a staggered transverse magnetic field as a function of $\arctan(\eta)$ between $-\pi/2$ and $\pi/2$. The plot corresponds to the blue horizontal line indicated by $\xi = 2J$ in Fig.~\ref{fig:ic-phase_diagram}. Real parts are depicted using solid lines, imaginary parts using dotted lines. The gap $\Delta = 2 \omega^-_0$ is depicted in black, the other elementary excitation with the same momentum $2 \omega^+_0$ in gray. The energies are complex conjugated. Their real parts vanish at $\eta=0$, indicating a second-order phase transition.}
		\label{fig:ic-gap-line}
	\end{figure}
	%%%%%%%%%%%%%%%%%%%%%%%%%%%%%%%%%%%%%%%%%%%%%%%%%%%%%%%%%%%%%%%%%%%%%%%%%%%%%%%%%%%%%%%%%%%%
	
	The gap $\Delta = 2 \omega^-_0$ is one of the energies that are real in certain extended regions of the parameter space, namely the region defined by $\xi \leq J$. Outside of this region it is complex conjugated to $2 \omega^+_0$, the other elementary excitation with the same momentum $k=0$. More specifically, it is given by
	\begin{align}
		\omega^\pm_0 = \begin{cases}
			\left| \eta \pm \sqrt{J^2 - \xi^2} \right|, & \xi \leq J\\
			\eta \pm \imag \sqrt{\xi^2 - J^2}, & \xi \geq J.
		\end{cases}
	\end{align}

	The closing behavior of this gap indicates the same phase transitions as indicated by the ground-state energy. At $r = 1$ we have $\Delta = 0$ as in the Hermitian limit. This is depicted in \mbox{Fig.~\ref{fig:ic-gap-radial}} for different $\varphi$. At $\eta = 0$ we have
	\begin{align}
		\Delta\big|_{\eta = 0} = 2 \sqrt{J^2 - \xi^2},
	\end{align}
	which has a vanishing real part for $\xi > J$, but a non-vanishing imaginary part. This is depicted in Fig.~\ref{fig:ic-gap-line}. Since complex eigenvalues of $PT$-symmetric Hamiltonians come in conjugate pairs, these imaginary parts cancel in the ground-state energy and the vanishing real part plays the part of a vanishing gap. 
	
As a consequence, the gap vanishes with an exponent $z\nu=1/2$ for $\eta = 0$. In all other situations with $\eta \neq 0$ the gap closes linearly with $z\nu=1$ including the well known Hermitian case where the gap is exactly given by the first-order expression $|J-\eta|$.
	
Now we have that the ground-state energy is real in the whole parameter space and the gap in the region defined by $|\xi| \leq J$. The question arises whether we can find a region with a purely real spectrum therein. This question is answered quickly by observing that e.g. $\omega^-_\pi$ is only real on the Hermitian axis defined by $\xi = 0$. Therefore we can conclude that there exist states with broken $PT$-symmetry as soon as we turn on the non-Hermitian perturbation.

%Ising chain - Results with series expansions
%%%%%%%%%%%%%%%%%%%%%%%%%%%%%%%%%%%%%%%%%%%%%%%%%%%%%%%%%%%%%%%%%%%%%%%%%%%%%%%%%%%%%%%%%%%%
\subsection{Gauging high-order series expansions}

We can now use the analytic solution to gauge our series expansion results. We derived the gap up to \nth{10} order about the low-field limit \eqref{eq:ic-Low_Field-Gap} and about the high-field limit \eqref{eq:ic-High_Field-Gap}. This order is similar to the maximal order for the 2D case discussed below where the exact solution is not available. Note that in principle we can reach much higher orders for the non-Hermitian Ising chain by simply determining the Taylor series of the analytic expression.
	
	%Figure - Ising chain - Gap - Approximants
	%%%%%%%%%%%%%%%%%%%%%%%%%%%%%%%%%%%%%%%%%%%%%%%%%%%%%%%%%%%%%%%%%%%%%%%%%%%%%%%%%%%%%%%%%%%%
	\begin{figure}[t]
		\centering
		\resizebox{\columnwidth}{!}{\input{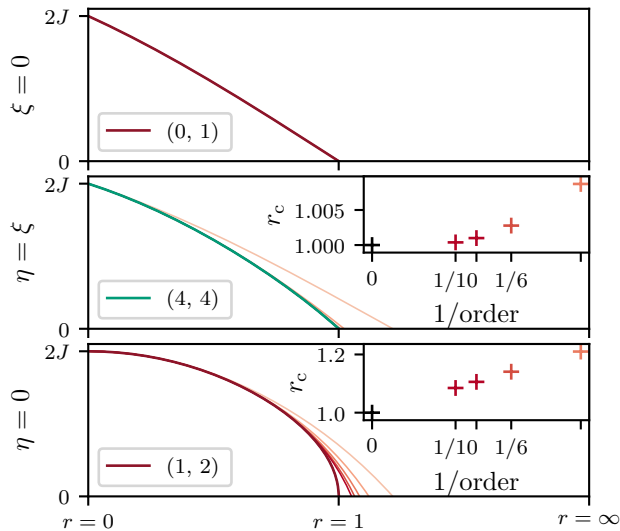}}
		\caption{Real part of the gap for the non-Hermitian Ising chain in a staggered transverse magnetic field as a function of $\arctan(r)$ between $0$ and $\pi/2$. The three plots correspond to the three yellow radial paths with $\xi=0$ (upper panel), $\eta=\xi$ (middle panel), and $\eta=0$ (lower panel) in Fig.~\ref{fig:ic-phase_diagram}. The analytic solution is depicted in black; the series expansions (except for the \nth{0} order) up to \nth{10} order in red. Selected DLog Padé approximants \eqref{eq:Integrated-Pade} are plotted in thick lines that are colored according to the colors in Fig.~\ref{fig:ic-DLog-Pade-Diagonal-lf}. The roots of the series expansion (up to orders 4, 6, 8, 10 from right to left) are plotted as a function of $1/\textrm{order}$ in inset plots in red. The black markers indicate the critical point $\rc = 1$ in both cases.}
		\label{fig:ic-gap-approximants-real}
	\end{figure}
	%%%%%%%%%%%%%%%%%%%%%%%%%%%%%%%%%%%%%%%%%%%%%%%%%%%%%%%%%%%%%%%%%%%%%%%%%%%%%%%%%%%%%%%%%%%%
	
	Except for $\eta = 0$ and $\xi = 0$, the vanishing of the gap is only accessible from the low-field limit due to the presence of an exceptional point at $r>\rc$ as can be seen exemplary in the middle panel of FIG.~\ref{fig:ic-gap-radial}. This is different for $\eta = 0$, where the two points coincide. The analytic low-field gap, as well as its series expansion order by order, is plotted in Fig.~\ref{fig:ic-gap-approximants-real} for exemplary regions of the parameter space. The Hermitian case $\xi = 0$ is exact already in first order, all other cases are only approximated by our expansions. The critical point is predicted more accurately the closer we are to the Hermitian limit. Using the bare series in tenth order, it is indicated at $\rc = 1.0004$ for $\eta = \xi$ and at $\rc = 1.08$ for $\eta = 0$.
	
	%Figure - Ising chain - Gap (eta = xi)
	%%%%%%%%%%%%%%%%%%%%%%%%%%%%%%%%%%%%%%%%%%%%%%%%%%%%%%%%%%%%%%%%%%%%%%%%%%%%%%%%%%%%%%%%%%%%
	\begin{figure}[t]
		\centering
		\resizebox{\columnwidth}{!}{\input{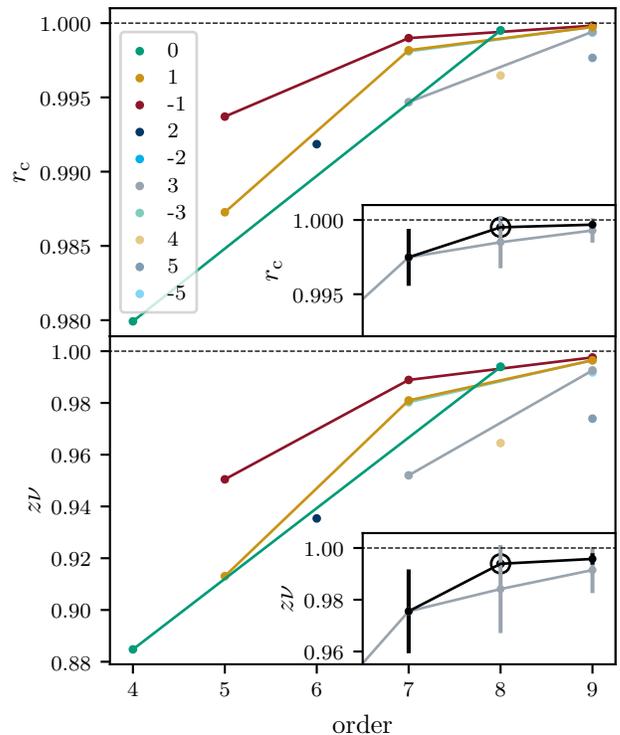}}
		\caption{Critical point $\rc$ (upper panel) and critical exponent $z \nu$ (lower panel) of the non-Hermitian Ising chain in a staggered transverse magnetic field as indicated by DLog Padé approximants. The parameter region is defined by $\eta = \xi$. Every family has a color assigned and is labeled by their difference between numerator and denominator degree. The averaged values and the respective sample standard deviations are displayed in the inset plots. The gray lines are obtained by including single family members, the black ones by considering them defective. In the latter case, there is only one approximant left in \nth{8} order, which is indicated by a circle. In \nth{9} order, the critical point [critical exponent] is found at \mbox{$\rc = 0.9993(8)$} [$z \nu = 0.991(9)$] when including single family members and at $\rc = 0.99967(19)$ [$ z \nu= 0.9955(23)$] when considering them defective. The dashed black lines indicate the analytical findings $\rc = 1$  and $z \nu = 1$.}
		\label{fig:ic-DLog-Pade-Diagonal-lf}
	\end{figure}
	%%%%%%%%%%%%%%%%%%%%%%%%%%%%%%%%%%%%%%%%%%%%%%%%%%%%%%%%%%%%%%%%%%%%%%%%%%%%%%%%%%%%%%%%%%%%
	
	The values are expected to become more accurate upon using DLog Padé approximants. With Eq.~\eqref{eq:Integrated-Pade}, selected approximants are depicted in Fig.~\ref{fig:ic-gap-approximants-real} as well. Remarkably, not only the limiting case $\xi = 0$ is described exactly by these approximants, but also the purely imaginary case $\eta = 0$ where the gap closes as a square root. This is similar to the extrapolation of the mean-field series in the limit of large spatial dimensions for the transverse-field Ising model on the hypercubic lattice \cite{Coester2016}. In fact, one can construct $\Delta = 2\sqrt{J^2 - \xi^2} - 2\eta$ exactly by adding the exact extrapolations on the real and imaginary axis. The same holds for the imaginary part of the gap about the high-field limit.
	
	For any given order, there are multiple different DLog Padé approximants to consider. For $\eta = \xi$, their respective critical points are plotted in the upper panel of Fig.~\ref{fig:ic-DLog-Pade-Diagonal-lf}. Per order the predicted values are obtained by averaging over all results by non-defective approximants. If single family members are considered defective, the predicted value for the critical point is given by \mbox{$\rc = 0.99967(19)$}, if they are included it is given by $\rc = 0.9993(8)$. Both values are accurate to the same order of magnitude as the value indicated by the bare \nth{10} order series expansion $\rc = 1.0004$.
	DLog Padé approximants are especially well suited to obtain critical exponents. For $\eta = \xi$, the critical exponent $z \nu$ is plotted in the lower panel of Fig.~\ref{fig:ic-DLog-Pade-Diagonal-lf}. If single family members are considered defective, the predicted value is given by $z \nu = 0.9955(23)$, if they are included it is given by \mbox{$z \nu = 0.991(9)$}. The accuracy is lower than the accuracy of the predicted critical point, which is expected. Nevertheless, it is still convincing.

	%Toric code - Results
	%%%%%%%%%%%%%%%%%%%%%%%%%%%%%%%%%%%%%%%%%%%%%%%%%%%%%%%%%%%%%%%%%%%%%%%%%%%%%%%%%%%%%%%%%%%%
	\section{Toric code results}\label{sec::tc}
	
	%Figure - Toric code - Gap - Approximants
	%%%%%%%%%%%%%%%%%%%%%%%%%%%%%%%%%%%%%%%%%%%%%%%%%%%%%%%%%%%%%%%%%%%%%%%%%%%%%%%%%%%%%%%%%%%%
	\begin{figure}[t]
		\centering
		\resizebox{\columnwidth}{!}{\input{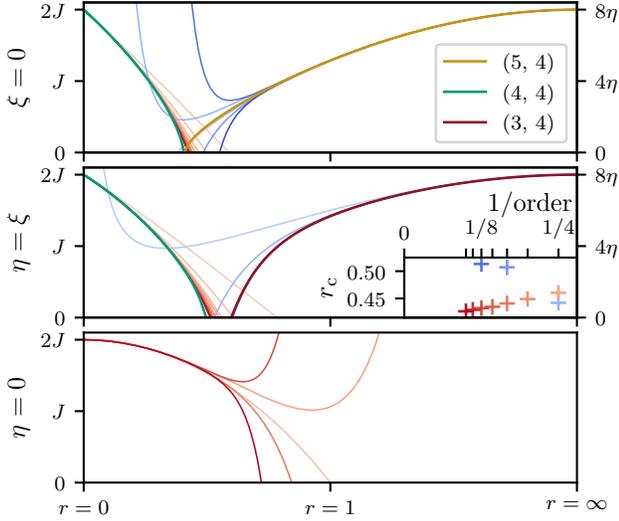}}
		\caption{Gap of the non-Hermitian toric code in a staggered parallel magnetic field as a function of $\arctan(r)$ between $0$ and $\pi/2$. The three plots correspond to the three different parameter regions with $\xi=0$ (upper panel), $\eta=\xi$ (middle panel), and $\eta=0$ (lower panel). The series expansions (except for the \nth{0} order) up to \nth{10} order are depicted in red around the low-field limit and in blue around the high-field limit, with larger orders having darker colors. The high-field expansion for \mbox{$\eta = \xi$} is only available up to \nth{8} order. Selected DLog Padé approximants \eqref{eq:Integrated-Pade} are plotted using thick lines with colors matching the respective families. The roots of the high-field series expansion for $\eta = \xi$ (up to orders 4, 6, 8 from right to left) are plotted as a function of $1/\textrm{order}$ in the inset plot in blue, those of the respective low-field expansion (up to orders 4, 5, 6, 7, 8, 9, 10 from right to left) in red. %Note that the plots have two different scales. The low-field expansions scale like the left axis and start at $2J$, the large-field expansions scale like the right axis and start at $8\eta$.}
		}
		\label{fig:tc-gap-approximants}
	\end{figure}
	%%%%%%%%%%%%%%%%%%%%%%%%%%%%%%%%%%%%%%%%%%%%%%%%%%%%%%%%%%%%%%%%%%%%%%%%%%%%%%%%%%%%%%%%%%%%
	
	In contrast to the non-Hermitian Ising chain, there is no analytic solution for the toric code in a non-Hermitian staggered field whose low-energy physics corresponds to the dual non-Hermitian transverse-field Ising model on the square lattice with non-Hermitian Ising interactions. However, the Hermitian case is well studied by different techniques including high-order series expansions \cite{Trebst_2007,Vidal_2009,Tupitsyn_2010,Dusuel_2011} (for the dual transverse-field Ising model on the square lattice see \cite{He_1990,Weihong_1994,Bloete2002,Kos2016}). One finds a second-order phase transition in the 3D* \cite{Schuler_2016} universality class with $\rc = 0.32847(4)$ \cite{Bloete2002} and $z \nu = 0.629971(4)$ \cite{Kos2016}. Here we are mainly interested in how the non-Hermitian staggered field influence this quantum critical breakdown of the topological phase. To this end we use the series expansion of the gap. The expansion about the low-field limit is given up to \nth{10} order in Eq.~\eqref{eq:tcg-Low_Field-Gap}. The expansion about the high-field limit \eqref{eq:tcg-High_Field-Gap} is given up to \nth{10} order in Eq.~\eqref{eq:tcg-High_Field-Gap-Hermitian} for $\xi = 0$ and up to \nth{8} order in Eq.~\eqref{eq:tcg-High_Field-Gap-Diagonal} for $\eta = \xi$. We note that the series in the Hermitian limits agree with the ones known from the literature \cite{He_1990,Weihong_1994}. All of these series expansions are plotted in Fig.~\ref{fig:tc-gap-approximants} order by order together with some selected DLog Padé approximants about both limits.

	The Hermitian case $\xi = 0$ is well studied by series expansions. We nevertheless summarize the main findings with the current perturbative order which allows a good comparison to the non-Hermitian case. DLog Padé approximation of the low-field gap results in \nth{8} order in $\rc = 0.328905(19)$ if single family members are included [if they are considered defective one finds $\rc = 0.3288978(10)$]. Here, we use the \nth{8} order results as there is only one non-defective approximant in \nth{9} order. The accuracy is expected to even increase for higher orders, as the families converge towards the literature value \mbox{$\rc = 0.32847(4)$} \cite{Bloete2002}. Extracting the corresponding critical exponent of the gap closing yields $z \nu = 0.6471(4)$ [$z \nu = 0.64698(18)$] when including [excluding] single family members. Turning to the gap in the high-field limit, one observes that the gap series is alternating and that it consists only of even powers in $r^{-1}$, which is equivalent to a series in $r^{-2}$. As a consequence, only three of the DLog Padé approximants are not defective (only two in ninth order). Taking the average and sample standard deviation yields a critical point of $\rc = 0.33394(18)$ and a critical exponent of $z \nu = 0.6354(23)$ in \nth{9} order. The result for the critical point fits the literature value, even though it is less accurate than the low-field results (the better agreement for the critical exponent is taken as a coincidence). In conclusion, the high-order series expansion does therefore well capture the second-order quantum phase transition between the low-field topological and the high-field polarized phase in a quantitative fashion. This is in particular true for the low-field expansion while the accuracy of the high-field expansion is less good.

	%Figure - Toric code - lf Gap (eta = xi)
	%%%%%%%%%%%%%%%%%%%%%%%%%%%%%%%%%%%%%%%%%%%%%%%%%%%%%%%%%%%%%%%%%%%%%%%%%%%%%%%%%%%%%%%%%%%%
	\begin{figure}[t]
		\centering
		\resizebox{\columnwidth}{!}{\input{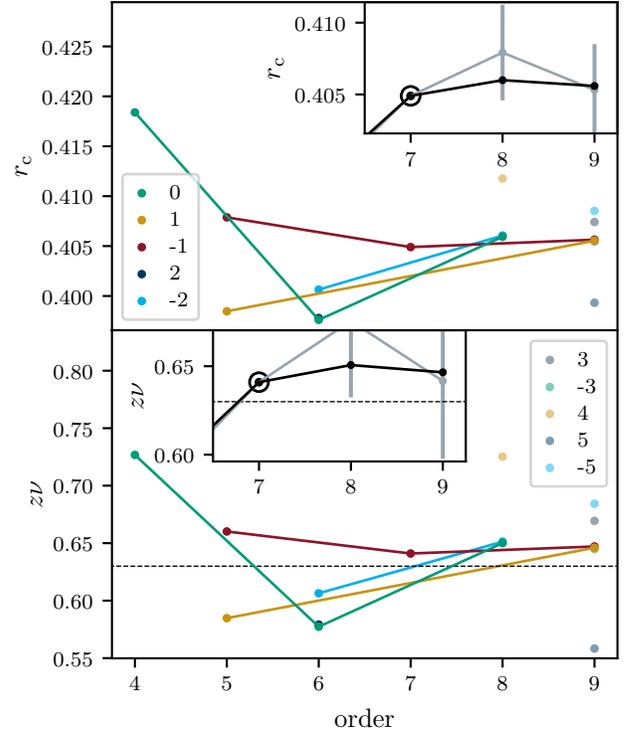}}
		\caption{Critical point $\rc$ (upper panel) and critical exponent $z\nu$ (lower panel) of the non-Hermitian toric code in a non-Hermitian staggered parallel magnetic field for $\eta = \xi$ obtained by DLog Padé approximants of the low-field gap as a function of the order. Every family of approximants has a distinct color and is labeled by their difference between numerator and denominator degree. The averaged values and the respective sample standard deviation are displayed in the inset plots. The gray line is obtained by including single family members, the black one by considering them defective. In the latter case, there is only one approximants left in \nth{7} order, which is indicated by a circle. In \nth{9} order, the critical point [exponent] is found at $\rc = 0.405(3)$ [$z\nu=0.64(4)$] when including single family members and $\rc = 0.40559(5)$ [$z\nu=0.6466(6)$] when considering them defective. The dashed black lines indicate the literature value $z \nu = 0.629971(4)$ \cite{Kos2016} for the exponent of the Hermitian model.}
		\label{fig:tc-DLog-Pade-Diagonal-lf}
	\end{figure}

	%Table - Toric code - hf Gap - Critical point (eta = xi)
	%%%%%%%%%%%%%%%%%%%%%%%%%%%%%%%%%%%%%%%%%%%%%%%%%%%%%%%%%%%%%%%%%%%%%%%%%%%%%%%%%%%%%%%%%%%%
	\begin{table}[t]
		\centering
		\begin{tabular}{c | l | l}
			DLog Padé & Critical point & Critical exponent\\ \hline
			$(3, 2)$ & $\rc = 0.38$ & $z \nu = 5.2$\\
			$(3, 4)$ & $\rc = 0.51$ & $z \nu = 1.1$\\
			$(5, 2)$ & $\rc = 0.58$ & $z \nu = 0.4$
		\end{tabular}
		\caption{Critical point and exponent of the non-Hermitian toric code in a staggered parallel magnetic field for $\eta = \xi$ as indicated by DLog Padé approximants about the high-field limit.}
		\label{tab:tc-DLog-Pade-Diagonal-hf}
	\end{table}
	%%%%%%%%%%%%%%%%%%%%%%%%%%%%%%%%%%%%%%%%%%%%%%%%%%%%%%%%%%%%%%%%%%%%%%%%%%%%%%%%%%%%%%%%%%%%
	%Table - Toric code - hf Gap - Critical point (eta = xi) - BIASED
	%%%%%%%%%%%%%%%%%%%%%%%%%%%%%%%%%%%%%%%%%%%%%%%%%%%%%%%%%%%%%%%%%%%%%%%%%%%%%%%%%%%%%%%%%%%%
	\begin{table}[t]
		\centering
		\begin{tabular}{c | l | l}
			Biased DLog Padé & Critical point & Prefactor\\ \hline
			$(3, 6)$ & $\rc = 0.5511$ & $a = 0.011453466$\\
			$(5, 4)$ & $\rc = 0.5460$ & $a = 0.00536516$\\
			$(7, 2)$ & $\rc = 0.5469$ & $a = 0.00609564$
		\end{tabular}
		\caption{Critical point and exponent of the non-Hermitian toric code in a staggered parallel magnetic field for $\eta = \xi$ as indicated by biased DLog Padé approximants about the high-field limit. The bias is taken such that the critical exponent is $z \nu = 0.629971$. The critical point is found at $\rc = 0.5480(23)$.}
		\label{tab:tc-Biased-DLog-Pade-Diagonal-hf}
	\end{table}
	%%%%%%%%%%%%%%%%%%%%%%%%%%%%%%%%%%%%%%%%%%%%%%%%%%%%%%%%%%%%%%%%%%%%%%%%%%%%%%%%%%%%%%%%%%%%
	Next we turn on the non-Hermitian staggered field and focus on $\eta = \xi$. Estimates of the critical point $\rc$ and critical exponent $z\nu$ obtained by DLog Padé approximation of the low-field gap are shown in Fig.~\ref{fig:tc-DLog-Pade-Diagonal-lf}. One finds $\rc = 0.405(3)$ with the highest order if single family members are included (at $\rc = 0.40559(5)$ if they are considered defective). The critical value is therefore shifted to larger values of $r$ in comparison to the Hermitian case. The series expansion about the high-field limit consists again only of even powers in $r^{-1}$. Except for the one corresponding to $r^{-2}$, all coefficients of this series are negative. The bare order 8 series indicates a critical point at $\rc \approx 0.51$. As for the Hermitian case, there are only three non-defective DLog Padé approximants; their respectively indicated critical points and exponents are listed in Tab.~\ref{tab:tc-DLog-Pade-Diagonal-hf}. Taking average and sample standard deviation yields a critical point of $\rc = 0.55(4)$ using the highest order. This result suggests that there could be at least two critical points in contrast to the Hermitian case. However, this result is to be handled with care, because the extrapolation of the high-field gap series does not work as well as the one for the low-field gap. This is also apparent because the critical exponents differ largely. We can, however, gain further insights by performing DLog Padé approximation biased with the 3D Ising critical exponent $z \nu = 0.629971(4)$, i.e., assuming no change in the universality class for finite non-Hermitian staggered fields. Although this must not be the correct exponent, the quality of the extrapolation is convincing. Because the series consists only of even powers in $r^{-1}$, the next higher order we add has to be even as well. More specifically, we add a term of the form $a r^{-10}$. The respective prefactors $a$ as well as corresponding critical points are stated in Tab.~\ref{tab:tc-Biased-DLog-Pade-Diagonal-hf}. The listed averaged critical point is given by $\rc = 0.5480(23)$. The critical points of the biased approximants match far better than those of the unbiased approximants, which supports our assumed critical exponent and the potential presence of an intermediate regime between the low-field and the high-field phase. If we instead perform a DLog Padé approximation biased with the critical point of the low-field expansion $\rc = 0.405$, i.e., assuming no intermediate phase, we obtain unphysical critical exponents. This further supports the potential presence of such an intermediate phase.

	%Table - Toric code - lf Gap - Critical point (eta = 0)
	%%%%%%%%%%%%%%%%%%%%%%%%%%%%%%%%%%%%%%%%%%%%%%%%%%%%%%%%%%%%%%%%%%%%%%%%%%%%%%%%%%%%%%%%%%%%
	\begin{table}[t]
		\centering
		\begin{tabular}{c | l | l}
			DLog Padé & Critical point & Critical exponent\\ \hline
			$(3, 2)$ & --- & ---\\
			$(3, 4)$ & $\rc = 1.17$ & $z \nu = 1.14$\\
			$(5, 2)$ & --- & ---\\
			$(3, 6)$ & $\rc = 0.89$ & $z \nu = 0.42$\\
			$(5, 4)$ & $\rc = 0.72$ & $z \nu = 0.12$\\
			$(7, 2)$ & --- & ---
		\end{tabular}
		\caption{Critical point and exponent of the non-Hermitian toric code in a staggered parallel magnetic field for $\eta = 0$ as indicated by DLog Padé approximants about the low-field limit. The approximants that do not close are indicated by dashes.}
		\label{tab:tc-DLog-Pade-Antihermitian-lf}
	\end{table}
	%%%%%%%%%%%%%%%%%%%%%%%%%%%%%%%%%%%%%%%%%%%%%%%%%%%%%%%%%%%%%%%%%%%%%%%%%%%%%%%%%%%%%%%%%%%%
	For the purely anti-Hermitian staggered field ($\eta = 0$), the series expansion of the low-field gap consists only of even powers in $r$ and is alternating. As a consequence, the extrapolation does not work in a convincing manner. Most of the DLog Padé approximants are defective and half of those that are not defective do not close. The associated critical points and exponents are listed in Tab.~\ref{tab:tc-DLog-Pade-Antihermitian-lf}. If one only considers the approximants that close, the critical point is indicated at $\rc = 0.80(9)$ using the highest order. If one considers all approximants, it is not certain that the gap closes at all. Furthermore, no large-field expansion is available for this case.
		
	%Figure - Toric code - Phase Diagram
	%%%%%%%%%%%%%%%%%%%%%%%%%%%%%%%%%%%%%%%%%%%%%%%%%%%%%%%%%%%%%%%%%%%%%%%%%%%%%%%%%%%%%%%%%%%%
	\begin{figure}[t]
		\centering
		\resizebox{\columnwidth}{!}{\input{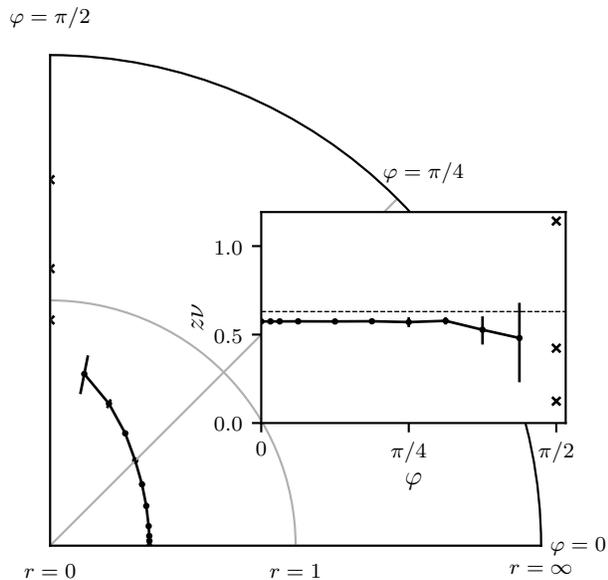}}
		\caption{Phase diagram of the non-Hermitian toric code in a staggered parallel magnetic field \eqref{eq:tc} obtained from DLog Padé approximants of the low-field gap. The radial plot displays the indicated critical point as a function of $\arctan(r)$ and $\varphi$ in polar coordinates. The inset plot displays the corresponding critical exponent as a function of $\varphi$. The points and errorbars are obtained by averaging over results by selected approximants and taking the corresponding sample standard deviation. For every point, selected are either all non-defective \nth{9} order approximants or, if there is only one in \nth{9} order, all non-defective \nth{8} order approximants. The critical point shifts to higher values of $r$ if the anti-Hermitian part increases. For large anti-Hermitian parts it is not certain that the gap closes at all. The displayed crosses are the values from Tab.~\ref{tab:tc-DLog-Pade-Antihermitian-lf}. The critical exponents are close to the 3D* Ising critical exponent $z \nu = 0.629971(4)$ \cite{Kos2016} that is indicated by the dashed line.}
		\label{fig:tc-phase_diagram}
	\end{figure}
	%%%%%%%%%%%%%%%%%%%%%%%%%%%%%%%%%%%%%%%%%%%%%%%%%%%%%%%%%%%%%%%%%%%%%%%%%%%%%%%%%%%%%%%%%%%%
	We finally discuss the extension of the topological phase in the full parameter space which is illustrated in Fig.~\ref{fig:tc-phase_diagram} using DLog Padé approximation of the low-field gap series. The latter works reliably except for the regime of small $\eta$ as already discussed for the purely anti-Hermitian staggered field. We observe that the critical point shifts to higher values of $r$ for increasing $\xi$. The robustness of the topological phase therefore increases continuously with the non-Hermiticity of the staggered field. The corresponding averaged critical exponents are illustrated in Fig.~\ref{fig:tc-phase_diagram} as well. As long as the approximants are reliable, the deduced critical exponents are almost independent of $\varphi$ and close to the 3D* Ising critical exponent $z \nu = 0.629971(4)$ \cite{Kos2016}. We therefore conjecture that the critical breakdown of the topological phase remains in the 3D* Ising universality class in this parameter regime and can be described by the condensation of anyonic quasi-particles.

	%Conclusions
	%%%%%%%%%%%%%%%%%%%%%%%%%%%%%%%%%%%%%%%%%%%%%%%%%%%%%%%%%%%%%%%%%%%%%%%%%%%%%%%%%%%%%%%%%%%%
	\section{Conclusions}\label{sec:conclusions}
	
	In this work we have extended high-order linked-cluster expansions like perturbative continuous unitary transformations and Takahashi perturbation theory to non-Hermitian quantum spin models with $PT$-symmetry. In practice, we have reached similar maximal perturbative orders for the ground-state energy per site and the energy gap of elementary excitations as for the more conventional Hermitian counterparts. Here we have considered the one-dimensional Ising chain and the two-dimensional toric code in a non-Hermitian staggered field. For both systems we have exploited dual descriptions in terms of non-Hermitian staggered Ising interactions in an uniform transverse field.
	
	For the one-dimensional non-Hermitian Ising chain we have demonstrated an alternative analytic solution in full agreement with the one given in Ref.~\onlinecite{Li2014}. Its ground-state phase diagram can be characterized by logarithmic singularities of the Laplacian of the ground-state energy and the gap closing of the low-field gap in the symmetry-broken phase. In particular, the gap closing of the low-field gap can be quantitatively described by extrapolating the high-order gap series.
		
	For the two-dimensional toric code in a non-Hermitian staggered field we study the quantum robustness of the topologically ordered phase by the gap closing of the low-field gap. We find that the well-known second-order quantum phase transition of the toric code in an uniform field extends into a large portion of the non-Hermitian parameter space. However, the series expansions become unreliable for a dominant anti-Hermitian field. Interestingly, the analysis of the high-field gap reveals the potential presence of an intermediate region, which, however, deserves further studies in the future. It would be further interesting to use high-order series expansions also for other perturbed non-Hermitian toric codes \cite{Guo_2020,Guo_2020b, Shackleton_2020} where the correspondence between bulk quasi-particles and topologically protected degenerate ground states breaks down. At this point we stress that we have concentrated here on applying high-order series expansions to extract spectral properties like the ground-state energy and elementary gaps in non-Hermitian quantum spin systems. Extensions towards other physical quantities are certainly possible and interesting. In particular, entanglement measures like the entanglement entropy would directly probe topological quantum order.
	
	%Acknowledgment
	%%%%%%%%%%%%%%%%%%%%%%%%%%%%%%%%%%%%%%%%%%%%%%%%%%%%%%%%%%%%%%%%%%%%%%%%%%%%%%%%%%%%%%%%%%%%
	\section*{Acknowledgements}
	
	KPS acknowledges financial support by the German Science Foundation (DFG) through the grant SCHM 2511/11-1. LL and KPS acknowledge support by the Deutsche Forschungsgemeinschaft (DFG, German Research Foundation) -- Project-ID 429529648 -- TRR 306 QuCoLiMa ("Quantum Cooperativity of Light and Matter’’).
	
	%Appendix
	%%%%%%%%%%%%%%%%%%%%%%%%%%%%%%%%%%%%%%%%%%%%%%%%%%%%%%%%%%%%%%%%%%%%%%%%%%%%%%%%%%%%%%%%%%%%
	\clearpage
	\onecolumngrid
	\appendix
	
	%Appendix - pCUT
	%%%%%%%%%%%%%%%%%%%%%%%%%%%%%%%%%%%%%%%%%%%%%%%%%%%%%%%%%%%%%%%%%%%%%%%%%%%%%%%%%%%%%%%%%%%%
	\section{Low-field expansions} \label{sec:Appendix-Low_Field}
	The ground-state energy per spin of the Ising chain in a non-Hermitian staggered transverse field \eqref{eq:ic} is given by
	\begin{align}
		e_0^{(12, \eta, \xi)} &= - J - \frac{1}{4} \frac{\eta^2 - \xi^2}{J} - \frac{1}{64} \frac{\eta^4 + \xi^4}{J^3} - \frac{5}{32} \frac{\eta^2 \xi^2}{J^3} - \frac{1}{256} \frac{\eta^6 - \xi^6}{J^5} + \frac{7}{256} \frac{\eta^4 \xi^2 - \eta^2 \xi^4}{J^5} - \frac{25}{16384} \frac{\eta^8 + \xi^8}{J^7} \nonumber\\
		&\hphantom{{}=} - \frac{129}{4096} \frac{\eta^6 \xi^2 + \eta^2 \xi^6}{J^7} - \frac{171}{8192} \frac{\eta^4 \xi^4}{J^7} - \frac{49}{65536} \frac{\eta^{10} - \xi^{10}}{J^9} + \frac{781}{65536} \frac{\eta^8 \xi^2 - \eta^2 \xi^8}{J^9} - \frac{33}{32768} \frac{\eta^6 \xi^4 - \eta^4 \xi^6}{J^9}  \label{eq:ic-Low_Field-Ground_State_Energy}\\
		&\hphantom{{}=} - \frac{441}{1048576} \frac{\eta^{12} + \xi^{12}}{J^{11}} - \frac{7631}{524288} \frac{\eta^{10} \xi^2 + \eta^2 \xi^{10}}{J^{11}} - \frac{18551}{1048576} \frac{\eta^8 \xi^4 + \eta^4 \xi^8}{J^{11}} - \frac{7241}{262144} \frac{\eta^6 \xi^6}{J^{11}}. \nonumber
	\end{align}
	The ground-state energy per spin of the toric code in a non-Hermitian staggered field \eqref{eq:tcg} is given by
	\begin{align}
		e_0^{(12, \eta, \xi)} &= - J - \frac{1}{4} \frac{\eta^2 - \xi^2}{J} - \frac{15}{64} \frac{\eta^4 + \xi^4}{J^3} - \frac{11}{32} \frac{\eta^2 \xi^2}{J^3} - \frac{147}{256} \frac{\eta^6 - \xi^6}{J^5} - \frac{131}{256} \frac{\eta^4 \xi^2 - \eta^2 \xi^4}{J^5} - \frac{18003}{8192} \frac{\eta^8 + \xi^8}{J^7} \label{eq:tcg-Low_Field-Ground_State_Energy}\\
		&\hphantom{{}=} - \frac{1699}{512} \frac{\eta^6 \xi^2 + \eta^2 \xi^6}{J^7} - \frac{9981}{4096} \frac{\eta^4 \xi^4}{J^7} - \frac{5420775}{524288} \frac{\eta^{10} - \xi^{10}}{J^9} - \frac{29368399}{1572864} \frac{\eta^8 \xi^2 - \eta^2 \xi^8}{J^9} - \frac{6703565}{786432} \frac{\eta^6 \xi^4 - \eta^4 \xi^6}{J^9} \nonumber\\
		&\hphantom{{}=} - \frac{33446240377}{603979776} \frac{\eta^{12} + \xi^{12}}{J^{11}} - \frac{108757489433}{905969664} \frac{\eta^{10} \xi^2 + \eta^2 \xi^{10}}{J^{11}} - \frac{6105915487}{67108864} \frac{\eta^8 \xi^4 + \eta^4 \xi^8}{J^{11}} - \frac{23451561631}{452984832} \frac{\eta^6 \xi^6}{J^{11}}. \nonumber
%		&\hphantom{{}=} - \frac{86903888779}{268435456} \frac{\eta^{14} - \xi^{14}}{J^{13}} - \frac{5911677814373}{7247757312} \frac{\eta^{12} \xi^2 - \eta^2 \xi^{12}}{J^{13}} - \frac{16122874716715}{21743271936} \frac{\eta^{10} \xi^4 - \eta^4 \xi^{10}}{J^{13}} \nonumber\\
%		&\hphantom{{}=} - \frac{5403932804599}{21743271936} \frac{\eta^8 \xi^6 - \eta^6 \xi^8}{J^{13}}. \label{eq:tcg-Low_Field-Ground_State_Energy}
	\end{align}
	The gap of the Ising chain in a non-Hermitian staggered transverse field \eqref{eq:ic} is given by
	\begin{align}
		\Delta^{(10, \eta, \xi)} &= 2 J - 2 \eta - \frac{\xi^2}{J} - \frac{1}{4} \frac{\xi^4}{J^3} - \frac{1}{8} \frac{\xi^6}{J^5} - \frac{5}{64} \frac{\xi^8}{J^7} - \frac{7}{128} \frac{\xi^{10}}{J^9}. \label{eq:ic-Low_Field-Gap}
	\end{align}
	The gap of the toric code in a non-Hermitian staggered field \eqref{eq:tcg} is given by
	\begin{align}
		\Delta^{(10,\eta,\xi)} &= 2 J - 4 \eta - 2 \frac{\eta^2}{J} - 2 \frac{\xi^2}{J} - 3 \frac{\eta^3}{J^2} - 3 \frac{\eta \xi^2}{J^2} - \frac{9}{2} \frac{\eta^4}{J^3} - 3 \frac{\eta^2 \xi^2}{J^3} + \frac{3}{2} \frac{\xi^4}{J^3} - 11 \frac{\eta^5}{J^4} - \frac{23}{2} \frac{\eta^3 \xi^2}{J^4} - \frac{1}{2} \frac{\eta \xi^4}{J^4} - \frac{2625}{128} \frac{\eta^6}{J^5} \nonumber\\
		&\hphantom{{}=} - \frac{3325}{128} \frac{\eta^4 \xi^2}{J^5} - \frac{1463}{128} \frac{\eta^2 \xi^4}{J^5} - \frac{763}{128} \frac{\xi^6}{J^5} - \frac{14771}{256} \frac{\eta^7}{J^6} - \frac{21595}{256} \frac{\eta^5 \xi^2}{J^6} - \frac{8733}{256} \frac{\eta^3 \xi^4}{J^6} - \frac{1909}{256} \frac{\eta \xi^6}{J^6} - \frac{940739}{8192} \frac{\eta^8}{J^7} \nonumber\\
		&\hphantom{{}=} - \frac{180361}{1024} \frac{\eta^6 \xi^2}{J^7} - \frac{227827}{4096} \frac{\eta^4 \xi^4}{J^7} + \frac{557}{16} \frac{\eta^2 \xi^6}{J^7} + \frac{238689}{8192} \frac{\xi^8}{J^7} - \frac{11472297}{32768} \frac{\eta^9}{J^8} - \frac{1274661}{2048} \frac{\eta^7 \xi^2}{J^8} - \frac{4894637}{16384} \frac{\eta^5 \xi^4}{J^8} \nonumber\\
		&\hphantom{{}=} + \frac{55859}{4096} \frac{\eta^3 \xi^6}{J^8} + \frac{1313867}{32768} \frac{\eta \xi^8}{J^8} - \frac{287258435}{393216} \frac{\eta^{10}}{J^9} - \frac{183060165}{131072} \frac{\eta^8 \xi^2}{J^9} - \frac{179805973}{196608} \frac{\eta^6 \xi^4}{J^9} - \frac{86807677}{196608} \frac{\eta^4 \xi^6}{J^9} \nonumber\\
		&\hphantom{{}=} - \frac{47455265}{131072} \frac{\eta^2 \xi^8}{J^9} - \frac{66440327}{393216} \frac{\xi^{10}}{J^9}. \label{eq:tcg-Low_Field-Gap}
	\end{align}
	
	%Appendix - Takahashi
	%%%%%%%%%%%%%%%%%%%%%%%%%%%%%%%%%%%%%%%%%%%%%%%%%%%%%%%%%%%%%%%%%%%%%%%%%%%%%%%%%%%%%%%%%%%%
	\section{High-field expansions} \label{sec:Appendix-High_Field}
	
	The ground-state energy per spin of the Ising chain in a non-Hermitian staggered transverse field \eqref{eq:ic} is given by
	\begin{align}
		\begin{split}
			e_0^{(10, J)} &= - \eta - \frac{J^2}{4 \eta} - \frac{J^4}{64 \eta^3} \frac{\eta^2 - 3 \xi^2}{\eta^2 + \xi^2} - \frac{J^6}{256 \eta^5} \frac{\eta^2 + 5 \xi^2}{\eta^2 + \xi^2} - \frac{J^8}{16384 \eta^7} \frac{25 \eta^6 - 269 \eta^4 \xi^2 - 405 \eta^2 \xi^4 - 175 \xi^6}{(\eta^2 + \xi^2)^3}\\
			&\hphantom{{}=} - \frac{J^{10}}{65536 \eta^9} \frac{49 \eta^6 + 715 \eta^4 \xi^2 + 1043 \eta^2 \xi^4 + 441 \xi^6}{(\eta^2 + \xi^2)^3}. \label{eq:ic-High_Field-Ground_State_Energy}
		\end{split}
	\end{align}
	The ground-state energy per spin of the toric code in a non-Hermitian staggered field \eqref{eq:tcg} is given by
	\begin{align}
		e_0^{(6, J)} &= - \eta - \frac{J}{2} - \frac{J^2}{16 \eta} - \frac{J^4}{1024 \eta^3} \frac{3 \eta^2 - 5 \xi^2}{9 \eta^2 + \xi^2} - \frac{J^6}{32768 \eta^5} \frac{54 \eta^6 + 9 \eta^4 \xi^2 + 368 \eta^2 \xi^4 + 29 \xi^6}{(9 \eta^2 + \xi^2)^2 (4 \eta^2 + \xi^2)}. \label{eq:tcg-High_Field-Ground_State_Energy}
	\end{align}
	The energy of the elementary excitations with $k = 0$ of the Ising chain in a non-Hermitian staggered transverse field \eqref{eq:ic} is given by
	\begin{align}
		\Delta_\pm^{(10, J)} &= 2 (\eta \pm \textrm{i} \xi) \pm \frac{1}{\textrm{i} \xi} J^2 \pm \frac{1}{4 \textrm{i} \xi^3} J^4 \pm \frac{1}{8 \textrm{i} \xi^5} J^6 \pm \frac{5}{64 \textrm{i} \xi^7} J^8 \pm \frac{7}{128 \textrm{i} \xi^9} J^{10} \label{eq:ic-High_Field-Gap}
	\end{align}
	for $\xi \neq 0$. The gap of the toric code in a non-Hermitian staggered field \eqref{eq:tcg} is given by
	\begin{align}
		\Delta^{(4, J)} = \begin{cases}
			\displaystyle 8 \eta + \frac{J^2}{4 \eta} \frac{- 3 \eta^2 + 5 \xi^2}{\eta^2 + \xi^2} + \frac{J^4}{768 \eta^3} \frac{387 \eta^8 - 5232 \eta^6 \xi^2 - 258 \eta^4 \xi^4 - 936 \eta^2 \xi^6 - 153 \xi^8}{(\eta^2 + \xi^2)^3 (9 \eta^2 + \xi^2)}, & \eta \geq \xi\\
			\displaystyle 8 \eta + \frac{J^2}{4 \eta} + \frac{J^4}{768 \eta^3} \frac{- 45 \eta^8 - 96 \eta^6 \xi^2 - 114 \eta^4 \xi^4 - 120 \eta^2 \xi^6 - 57 \xi^8}{(\eta^2 + \xi^2)^3 (9 \eta^2 + \xi^2)}, & \eta < \xi.
		\end{cases}\label{eq:tcg-High_Field-Gap}
	\end{align}
	For $\xi = 0$, this gap is given up to higher orders by
	\begin{align}
		\Delta^{(10,J)}\big|_{\xi = 0} = 8 \eta - \frac{3}{4} \frac{J^2}{\eta} + \frac{43}{768} \frac{J^4}{\eta^3} - \frac{19993}{884736} \frac{J^6}{\eta^5} + \frac{82873487}{10192158720} \frac{J^8}{\eta^7} - \frac{1901437203257}{587068342272000} \frac{J^{10}}{\eta^9} \label{eq:tcg-High_Field-Gap-Hermitian}% + \frac{64764934458802909}{47341191120814080000} \frac{J^{12}}{\eta^{11}}
	\end{align}
	and for $\eta = \xi$ by
	\begin{align}
		\Delta^{(8, J)}\big|_{\eta = \xi} = 8 \eta + \frac{1}{4} \frac{J^2}{\eta} - \frac{129}{1280} \frac{J^4}{\eta^3} - \frac{100511}{12288000} \frac{J^6}{\eta^5} - \frac{7866337483}{52140441600000} \frac{J^8}{\eta^7}. \label{eq:tcg-High_Field-Gap-Diagonal}
	\end{align}

	\twocolumngrid

\end{document}